# SELECTIVE WATCHDOG TECHNIQUE FOR INTRUSION DETECTION IN MOBILE AD-HOC NETWORK

## DEEPIKA DUA[1] AND ATUL MISHRA[2]

[1]Department of Computer Engineering, YMCA University of Science & Technology, Haryana , India
deepikadua876@gmail.com
[2]Department of Computer Engineering, YMCA University of Science & Technology, Haryana , India
mish.atul@gmail.com

### ABSTRACT

*Mobile ad-hoc networks(MANET) is the collection of mobile nodes which are self organizing and are connected by wireless links where nodes which are not in the direct range communicate with each other relying on the intermediate nodes. As a result of trusting other nodes in the route, a malicious node can easily compromise the security of the network. A black-hole node is the malicious node which drops the entire packet coming to it and always shows the fresh route to the destination, even if the route to destination doesn't exist. This paper describes a scheme that will detect the intrusion in the network in the presence of black-hole node and its performance is compared with the previous technique. This novel technique helps to increase the network performance by reducing the overhead in the network.*

## 1. INTRODUCTION

Mobile ad-hoc network is formed by the collection of some mobile nodes which can act both as a sender as well as receiver for data communication. They are decentralized networks which are self organizing and self maintaining. There is no fixed infrastructure in the network, the topology changes dynamically [1]. As a result of continuously changing topology, there is no fixed boundary of the network. The nodes cooperate with each other to forward the data packet. In such a network where there is no well-defined boundary, open medium, nodes rely on one other to forward the data packet, firewalls cannot be applied for securing these networks. Intrusion detection system [2] is used in these networks to detect the misbehaviour in the network. Intrusion detection system acts as a second layer in mobile ad-hoc networks [3].

In this paper, a scheme is proposed which detects the misbehaving nodes in the network in the presence of black-hole attack [4] [5] and reduces the network overhead.

Rest of the paper is organized as follows. In section 2, literature survey is presented. In section 3, scheme description is present, the methodology used is described. In section 4, simulation environment and results of the simulation are presented. And finally conclusion is presented in section5.

## 2. Related work

Marti el al [6] proposed a scheme named Watchdog which is a reputation based scheme [7], in which after detecting the malicious node, information is propagated throughout the network so to avoid that node in future routes.

The watchdog scheme works in two parts-in the first part the watchdog detects the malicious node by promiscuously listening to its next neighbour's transmission. If a node doesn't forward the packet after a threshold, then watchdog declares that node as malicious. And then the path rater finds the new route to the destination excluding that malicious node. In this scheme malicious node is detected instead of malicious link there are six weaknesses that are mentioned by Marti [1]. They are 1)Receiver Collision problem 2)Ambiguous collision 3)Limited Transmission power 4) False misbehaviour 5) collusion 6) Partial Dropping.

Liu at al [7] proposed a scheme named TWOACK, which detects the misbehaving links in the ad-hoc network instead of misbehaving nodes. It is an acknowledgement based scheme in which every third node in the route from sender to receiver requires to send an acknowledgement packet to the first node down the reverse route. It solves the receiver collision problem and limited power problem of the watchdog scheme. But due to the exchange of too many acknowledgement packets, this scheme consumes too much battery power and hence can degrade the network performance.

Sheltami et al. [8] proposed a scheme named Adaptive acknowledgment (AACK) which is based on TWOACK scheme. This scheme also works on DSR routing protocol. It is an advancement of the TWOACK scheme. It reduces the battery consumption by making the scheme a combination of end-to-end acknowledgement and TACK, which is similar to TWOACK. When the sender sends a data packet to destination, it waits for some time for the destination to acknowledge that data packet, but if the acknowledgement doesn't come within per-defined time, then it switches to TACK mode, where every third node sends the TACK packet to the nodes two hops away from it down the route.

Elhadi M. Shashuki, Nan Kang and Tarek R. Sheltami [9] proposed an approach called EAACK (Enhanced AACK) which solves receiver collisions problem, limited battery problem and false misbehaviour problem of the watchdog scheme. It is also an acknowledgement based scheme and to protect the acknowledgement packet from forging, this scheme makes use of digital signature. It is composed of three parts:-

ACK- It is an end-to-end acknowledgement as described in AACK scheme. Sender waits for the destination to acknowledge data packets but if the acknowledgement doesn't come within a specified time, then it switches to S-ACK mode.

S-ACK- In this mode, similar to TWOACK scheme, consecutive node works in a group i.e. every third node sends an S-ACK packet to its first node which is in the reverse directions. The difference between the S-ACK and TWOACK is that TWOACK immediately trusts the misbehaviour report and declares the node as malicious. But in this scheme, we switch to MRA mode to confirm the misbehaviour report.

MRA- It stands for misbehaviour report authentication. This mode cooperates with the routing protocol to find a new route to the destination which excludes the reported misbehaviour node. Destination is checked for the data packet using the new route. If the data packet is found at the destination, then it is a false misbehaviour report and the node which generated this report will be declared as malicious, else the misbehaviour report is trusted and the node would be declared as misbehaving.

## 3. Proposed Approach

We proposed an algorithm that detects the intrusion in the presence of black hole node in the network. The proposed technique is an improvement over the Watchdog technique[1].In Watchdog each node continuously hears its next node transmission but in the proposed selective Watchdog technique only when the acknowledgment would not be received ,then IDS would start. Morever, in watchdog[1] technique all nodes monitor their neighbours but in proposed selective watchdog technique ,network of nodes are divided into clusters and only nodes in the cluster which have value greater than threshold monitor their neighbours. The pseudo code for the black-hole attack is shown in the algorithm. The input parameters for the algorithm are set of all the nodes, a threshold value which gets updated dynamically, source node, destination node and all the nodes which send the route reply to the source node.

The algorithm works as follows:-

The source waits for the destination to send acknowledgement to it after every $10^{th}$ packet. If source receives the acknowledgement, then there is no misbehaviour in the network and process continues as such. But if the destination fails to acknowledge the data packets for a time period, then IDS starts its functionality.

As in black-hole attack, there is a greater possibility that black-hole node will send the highest sequence number to the source in route reply. The proposed IDS algorithm maintains the list of all the nodes which send the route reply to the source with sequence number greater than the threshold value.

The IDS will be applied only on those nodes which are in the list maintained by ids.

Algorithm 1: Algorithm for detecting IDS

Input: Threshold_seq_no. Set_of_all_nodes; Set_of_nodes_who_sent_route_reply; source; destination.

---

1.  Begin
2.  If(pkt_received_by_dest==pkt_sent_by_source) then
3.         network does not shows any malicious behaviour
4.  Else if(pkt_received_by_dest < certain percentage of pkt sent by source over the network)
5.  {
6.     Then the network shows malicious behaviour and IDS is applied to detect malicious behaviour
7.  For(int i=0; i<no._of_nodes_who_sent_route_reply; i++)
8.  {
9.         If(seq_no[route_reply[Node]] > Threshold_seq_no) then
10. List. add(next[Node])
11. List. add(Node)
12. List. add(prev[Node])
13. result= **Segment_watchdog(List)**;
14. If(result==true)   then //(i.e. if malicious node is found)
15.              Exit;
16.   ENDIF
17.   Else
18.       Continue
19.   EndElse

20. }
21. }

Algorithm: **Segmented_Watchdog (List)**

1. BEGIN
2. Result=false
3. malicious= Null
4. Node1= list. get(0)
5. Node2= list. get(1)
6. Node3=list. get(2)
7. //Chk(Sent_pkt[Node2]);
8. If(sent_pkt[Node2] == Received_pkt_by_node2) THEN
9.     Monitor Node3
10. ENDIF
11. If(Sent_pkt_by_Node3==Received_pkt_by_Node3) THEN
12.   No malicious activity detected in this segment
13.     RETURN Result
14. END IF
15. Else if(Sent_pkt_by_Node1 < Received_pkt_by_Node1) THEN
16.      Malicious= Node1
17.      Result= True
18.       RETURN Result
19. END ELSEIF
20. Else if(Sent_pkt_by_Node2 < Received_pkt_by_Node2) THEN
21.      Malicious= Node2
22.      Result=True
23.      RETURN Result
24. END ELSEIF
25. Else if(Sent_pkt_by_Node1 < Received_pkt_by_Node1) THEN
26.       Malicious= Node1
27.       Result=True
28.      RETURN Result
29. END ELSEIF
30. END

The Flowchart of the technique is shown in figure1

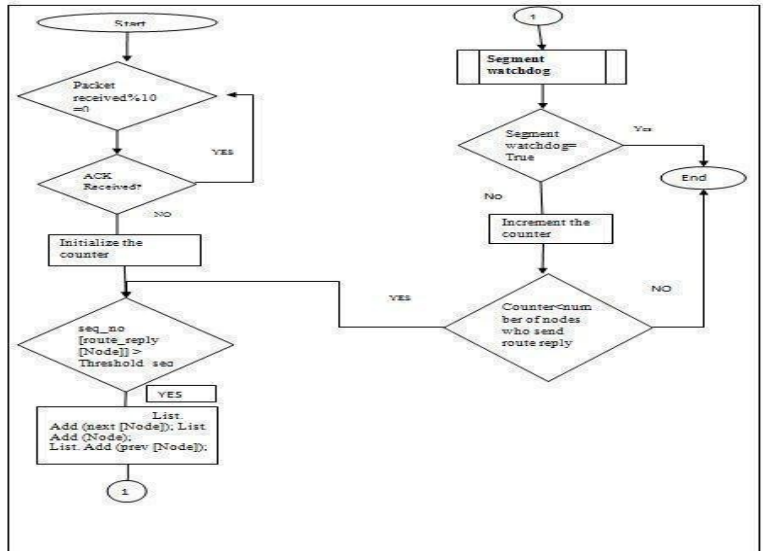

Figure 1(a). Flowchart of proposed technique

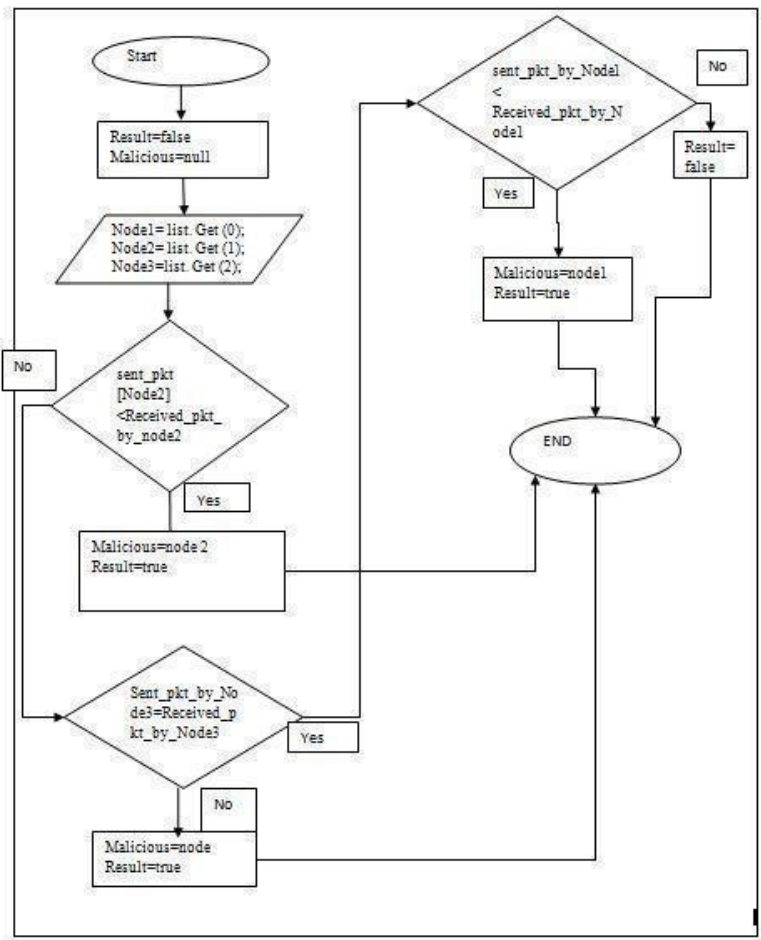

Figure 1(b). Flowchart of selective Watchdog

For every node in the list, segment watchdog method gets called. In this method, the number of packets send and received by the node is checked. If number of send and received are equal, then its successor node in the route is checked else its predecessor node in the route is evaluated in the same way.

## 4. Simulation Results

### 4.1. Assumptions

- We have assumed the bi-directionality in the links.
- Secondly, we have assumed that both the sender and receiver are trusted nodes, i.e. they are non-malicious.
- Duplicate MAC address doesn't exist.
- Lastly we have assumed that the nodes can overhear the transmission of their immediate neighbours.

### 4.2. Simulation Configuration

The Simulation is carried out using the tool Network Simulator 2 (NS-2) version 2.35 on Linux operating system Ubuntu version 12.10.The system runs on a laptop with Core 2 Duo T6500 processor with 4-GB RAM. For plotting graph, trace-graph version 202 is used.

- **Grid Size:** 500x500
- **Number of Nodes:** 10 of which 5 were communicating
- **Packet traffic:** CBR (Constant bit rate) on UDP
- **Packet Size:** 512B
- **Packet Interval:** 0.25
- **Routing Protocol:** AODV

### 4.3. Simulation Scenarios

To simulate our result we have taken two scenarios.
**Scenario 1**: In this scenario, Watchdog technique is implemented with one malicious node in path between source and destination.
**Scenario 2**: In this, Proposed Technique is implemented with same parameters taken in scenario 1.

### 4.4. Performance Evaluation

#### Scenario 1

In the first case, watchdog technique is implemented with a malicious node between the source and destination. Figure 2 shows the results that it detects the misbehaviour in the network of 10 nodes in 27.39 sec of neighbour detection.

```
deepi@ubuntu: /media/deepi/a/ns2/progs

detected neighbour pkt
        Node: IP 0 MAC 0
                Total  Rv:    0       Fw:    0       St:0

        Node: IP 2 MAC 2
                Total  Rv:    0       Fw:    0       St:0

        Node: IP 3 MAC 3
                Total  Rv:    111     Fw:    9       St:79
                UDP    Rv:    111     Fw:    9       St:79
                Node of 1 route/s: [0, 6],

        Node: IP 4 MAC 4
                Total  Rv:    0       Fw:    0       St:0

(Stored 4 neighbours)
----------
 new watchdog pkt
AlarmAlarm! node 3 (mac 3) not forward more than 20% packets: 20.54% loss for UD
P,  and 27.39 secs from neighbour detection
deepi@ubuntu:/media/deepi/a/ns2/progs$ 
```

Figure 2: Screenshot of Watchdog Technique

**Scenario 2**

In this scenario, proposed Selective Watchdog technique is implemented and then results are compared with the results of scenario 1.

Figure 3 show that it took 27.36 sec for our scheme to detect the intrusion in the network.

```
deepi@ubuntu: /media/deepi/a/ns2/progs
IP 0:0 (1) -> IP 6:0 (3), Type:2, Data: 111, Time: 27.402015

detected neighbour pkt
        Node: IP 1 MAC 1
                Total  Rv:    112     Fw:    111     St:1
                UDP    Rv:    112     Fw:    111     St:1
                Node of 1 route/s: [0, 6],

        Node: IP 0 MAC 0
                Total  Rv:    0       Fw:    0       St:0

        Node: IP 3 MAC 3
                Total  Rv:    110     Fw:    8       St:79
                UDP    Rv:    110     Fw:    8       St:79
                Node of 1 route/s: [0, 6],

(Stored 3 neighbours)
----------
 new watchdog pktpkt fwded deleting d pkt
helllooosss
Alarm! Alarm! node 3 (mac 3) not forward more than 20% packets: 20.72% loss for
UDP, 27.40 secs of execution and 27.36 secs from neighbour detection
deepi@ubuntu:/media/deepi/a/ns2/progs$ 
```

Figure 3: Screenshot of proposed technique

Figure 4 shows the graph of comparison between the proposed scheme and the watchdog scheme. From the graph, it is clearly shown that the proposed scheme performs better than the watchdog scheme in terms of detection time to detect the intrusion in the network.

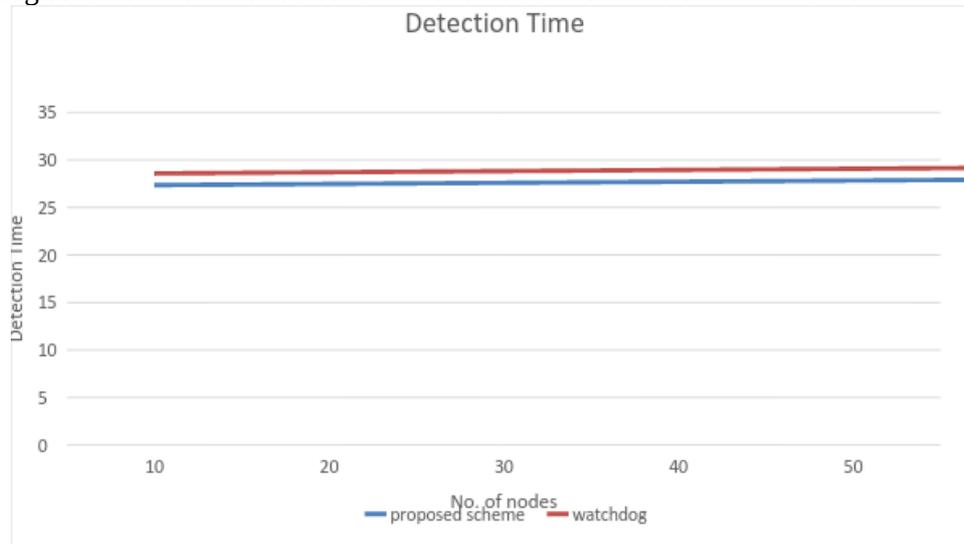

Figure4 Detection Time Comparison of proposed scheme and watchdog scheme

- **Quantitative Analysis**

For a network of n nodes, Watchdog scheme have n-2 promiscuous listening. As every node have to monitor its next neighbour except the source which is not monitored by any node and the destination which will not monitor any node?

For our proposed Selective Watchdog scheme, each cluster is of size say l, suppose we break the network of n nodes into K number of Clusters where K<<n i.e. n/l.
Let say a threshold value of T qualifies n/l*1/t, where t is a qualifier and its value will determine the number of clusters to be checked.
Promiscuous listening in a cluster of size l would be (l-2) in case both source and destination are included in the cluster and it would be l in other cases.
Total promiscuous listening in proposed study is l*(n\l-2) + 2*(l-2)
This formula calculates the number of promiscuous listening and it is for only one data packet.
Varying the number of cluster size and number of nodes taken, we can get different number of promiscuous listening.
Table 1 shows the different values taken using the above formula. For n=12,l=3,the number of promiscuous listening in Watchdog technique is 10 and in proposed technique is 8.Similarly,for other values shown in table , number of promiscuous listening is calculated.

Table 1.  Promiscuous listening with varying number of nodes and cluster size

|  | N=12 | N=24 | N=36 |
|---|---|---|---|
| L=3 | 8 | 20 | 32 |
| L=4 | 7 | 16 | 25 |
| L=6 | 8 | 14 | 20 |
| Watchdog Technique | 10 | 22 | 34 |

Figure 5 shows the graph for number of promiscuous listening for the proposed approach and the watchdog technique, plotted using the data provided in the table 1. The graph is plotted by varying the number of nodes and the size of cluster taken for each case.

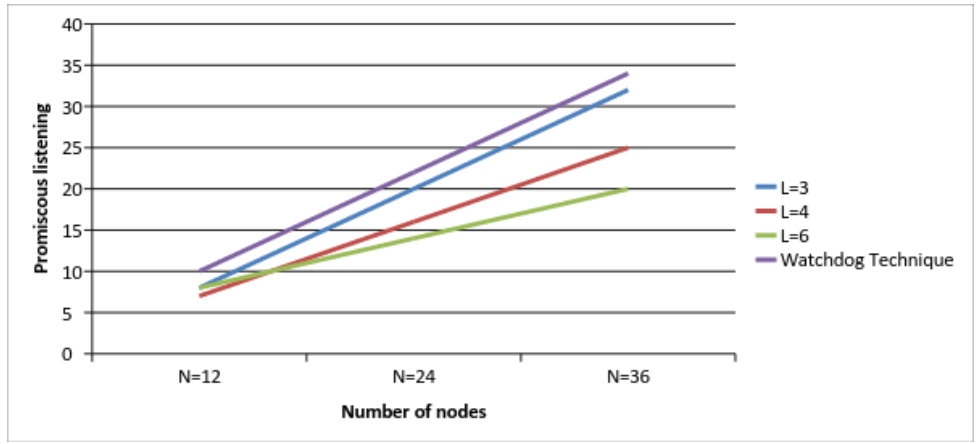

Figure 5 Number of nodes Vs Promiscuous Listening for proposed scheme

- **Experimental Analysis**

Number of nodes and size of cluster is varied and values are calculated by simulation. Table 2 shows the result of simulation.

Table 2 . Experimental value of promiscuous listening

|  | N 12 | N 24 | N 36 |
|---|---|---|---|
| L=3 | 238 | 580 | 900 |
| L=4 | 234 | 536 | 779 |
| L=6 | 220 | 448 | 589 |
| Watchdog Technique | 1109 | 2230 | 3689 |

From these values, the graph is plotted. Figure 6 show the graph plotted using the above values.

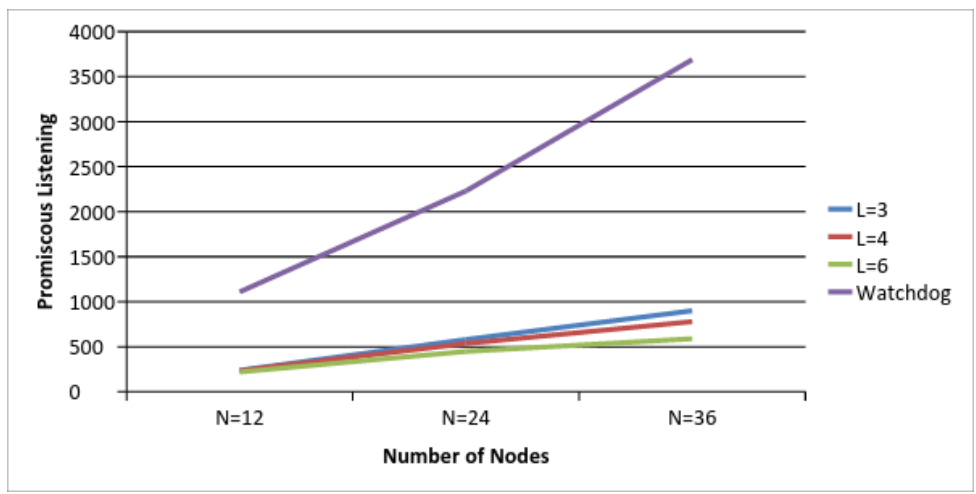

Figure 6 Number of nodes vs. Promiscuous listening

Table 3 shows the statistics of the number of packets sent, number of packets received and percentage of packets received and drop in all three scenarios i.e. in absence of malicious node, in its presence without IDS and with IDS. The statistics shows that with the presence of IDS in the network, the network performance gets improved.

Table 3: Statistics of simulation data

|  | Number of packets sent | Number of packets drop | Percentage of packets received | Percentage of packet drop |
|---|---|---|---|---|
| Absence of malicious node | 1000 | 3 | 99.69 | 0.30 |
| Presence of malicious node without IDS | 1000 | 997 | 0.30 | 99.7 |
| Watchdog Technique | 1000 | 120 | 88 | 12 |
| Proposed ID Technique | 1000 | 108 | 89.2 | 10.8 |

## 5. CONCLUSION

Security is the major concern in the ad-hoc networks as nodes can be easily captured or compromised. Black-hole attack drops all the packets coming to it. As a result network performance decrease drastically. The proposed scheme detects the intrusion in the presence of black-hole attack in the network and the results shows that it is better than Watchdog technique in terms of time to detect the intrusion and number of promiscuous listening. The graphs further shows that making the cluster and starting the IDS only when acknowledgment not received further improves the network throughput as there would be less network overhead.

**Authors**

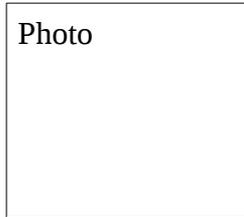

Short Biography